\documentstyle[onecolumn,psfig]{mn}

\begin{document}

\title{The Mass of Black Holes in the Nuclei of Spirals.}

\author[P. Salucci et al.]
{Paolo Salucci$^1$, Charu Ratnam$^2$, Pierluigi
Monaco$^3$ \& 
Luigi Danese$^1$ \\
 $^1$ International School for Advanced Studies, SISSA,
Via Beirut
2-4,
I-34013 Trieste, Italy\\ 
$^2$ Indian Institute of Astrophysics,Bangalore, India\\
$^3$ Institute of Astronomy, Madingley Road, CB3 0HA
Cambridge,
UK } 
\date{Received ; accepted }

\maketitle  

\markboth{BH masses in spirals}
{BH masses in Spirals} 

\begin{abstract}

We use the innermost  kinematics of spirals to investigate
whether these galaxies could host the massive Black Holes remnants
that once powered the QSO phenomenon. Hundreds rotation curves of 
early and late-type spirals are used to place upper limits on the
central BH masses.  We find: {\it i)} in late type
spirals,  the central massive dark  objects (MDO's)  are about
 $10-100$ times  smaller than the ones  detected in
ellipticals, {\it ii)} in  early type spirals,
 the central bodies are likely in the same mass range of the  elliptical MDO's.  
As a consequence, the contribution to the QSO/AGN
phenomenon by the BH remnants eventually hosted in Spirals is
negligible: $ \rho_{BH} (Sb-Im)<6 \times 10^4 M_\odot \ Mpc^{-3}$.
 We find  several
hints   that the MDO mass vs bulge mass relationship is significantly steeper in spirals 
than in ellipticals although, the  very issue of the existence 
of such relation for late Hubble type objects remains open.  
The upper limits on the mass of the BHs resident in late-type spirals
are stringent: $M_{BH}\leq 10^6 M_{\odot}-10^7 M_{\odot}$, indicating
that  only low luminosity activity possibly occurred in these objects.

\end{abstract}

\section {Introduction} 

There is an increasing direct evidence that, at the centers of
bulge-dominated galaxies, reside Massive Dark Objects (MDO), which are
the likely remnants of the engines that once powered the QSO
phenomenon (Ho, 1998). In a limited number of cases, the evidence
comes from the spectacular central-body-dominated (CBD) rotation
curves of circumnuclear ($0.1 \ pc - 2 \ pc$) gaseous disks obtained
by means of very high resolution instruments (i.e.  HST, VLBI and
VLBA, e.g. Bower et al. 1998, Cretton et al.  1998, Schreier et
al. 1998).

In general, however, the evidence is more indirect: only careful
studies of the dynamics of the innermost regions ($< 100 \ pc$) of
galaxies reveal the presence of a central compact object, allowing
mass estimates (e.g. Kormendy and Richstone 1995; Magorrian et al.\ 1998;
van der Marel 1997).  The conclusions are that virtually every hot
galaxy hosts a MDO with a mass ranging from $\sim 10^8 M_\odot$ to
$2\times 10^{10} M_\odot$ and roughly proportional to the mass of the
spheroidal stellar component ($M_b$). These
MDO masses are large enough to match those associated with the QSO
phenomenon. In fact, the highest bolometric luminosities of Quasars,
($ L_{bol} < 4 \times 10^{48} erg/s$), under the assumption that the
latter radiate at the Eddington limit, imply underlying BH masses of $ \sim 3 \times 10^{10} M_\odot$, i.e. masses
  comparable with those of the largest MDOs detected in ellipticals
(Magorrian et al.\ 1998). On the other hand,  the lowest QSO bolometric
luminosities $L_{bol}=10^{46} erg/s$ still imply considerable BH
masses,  of the order of $ 2 \times 10^8 M_\odot$ (e.g. Paper I) that it will
be hereafter considered as  the  minimum reference  mass for a BH
QSO-powering  remnant.   
For disk galaxies the situation is different and much more uncertain.
A direct determination of the central mass has been obtained only in a
few cases (including M31 and the Galaxy) and remarkably, these do not
exceed $10^7 M_\odot$ (see e.g. Paper I, Ho, 1998).  That is, in spirals, MDO masses in the
range of the QSO engine remnants have not currently been detected.
This could be an intrinsic property of galactic BH's, perhaps related
to the link between the BH mass and the bulge mass,  coupled with the
relationship between the latter and the host galaxy Hubble type.  In
any case, the eventual lack of detections in spirals of massive BH
($M_{BH} >10^8 M_\odot$) cannot be ascribed to observational biases.
In fact, the ``sphere of influence'' $r_{BH}$, of a black hole of mass
$M_{BH}$ at the center of a galaxy at distance $D$ and with a {\it
projected} central dispersion velocity $\sigma$ is given by: $
r_{BH}^{\prime \prime} \simeq 1 ^{\prime \prime} {M_{BH}\over 10^9\
M_\odot} \big({\sigma\over{ 200 km/s}}\big)^{-2 }\big({D\over {10
Mpc}} \big)^{-1}$ with $r_{BH}$ measured in arc-seconds.  In hot,
pressure-dominated systems, $\sigma$ ``balances'' the gravitational
attraction per unit mass due to a {\it great part} of the {\it whole}
galaxy: since $\sigma \sim 200 \ km/s $, the stellar dynamics, probed
at $1 ^{\prime \prime}$ resolution, can disentangle the central BH
from the surrounding galactic nucleus only if $M_{BH}\gg 10^8
M_\odot$.  In rotationally-dominated objects a relation like the above 
still approximately holds when we substitute $\sigma $ with the
circular velocity.  In this case, the  RCs (Rotation Curves), often  available
down to $R_{res}\sim 100 \ pc$,  if 
correctly probe   the gravitational potential, measure only  the mass {\it
inside } $R_{res}$. Therefore they  can resolve a central body of very small
mass in that  the    nuclear stellar mass, obtained by extrapolating at low
radii the stellar mass distribution, is  quite  small  $M(< 400 \ pc, stars)
\sim 10^{7-8}M_\odot  << M_{QSO}$ (e.g. PSS). Turning back the argument, if
nuclei of spirals host QSO remnants, then the innermost kpc of their  circular 
velocities will be strongly affected by such   central bodies.

In paper I (Salucci et al.  1998) we showed that the Mass Function of
Massive Dark Objects in elliptical galaxies matches that expected from
the remnants of the past QSO activity.  Extrapolating the E/S0 MDO
Mass Function, we have predicted that only low mass MDOs reside in
spirals and  that in these objects activity has been statistically lower
than in ellipticals.  The aim of this work is to confirm this 
scenario by investigating, through MDO's masses upper limits obtained
by inner kinematics, the predicted strong Hubble
Type dependence of the QSO remnant mass (Salucci et al. 1999, Monaco et al. 1999).
  
The plan of this work is the following: in section 2 we set up the
issue, in sections 3 and 4 we derive the MDO mass upper limits for
late and early type spirals, in section 4 we investigate the BH mass
vs bulge mass relationship for disk systems and in section 5 we derive
the constraints for the cosmological mass function of spiral BH.

\begin{figure}
\vspace{7.9cm}
\includegraphics{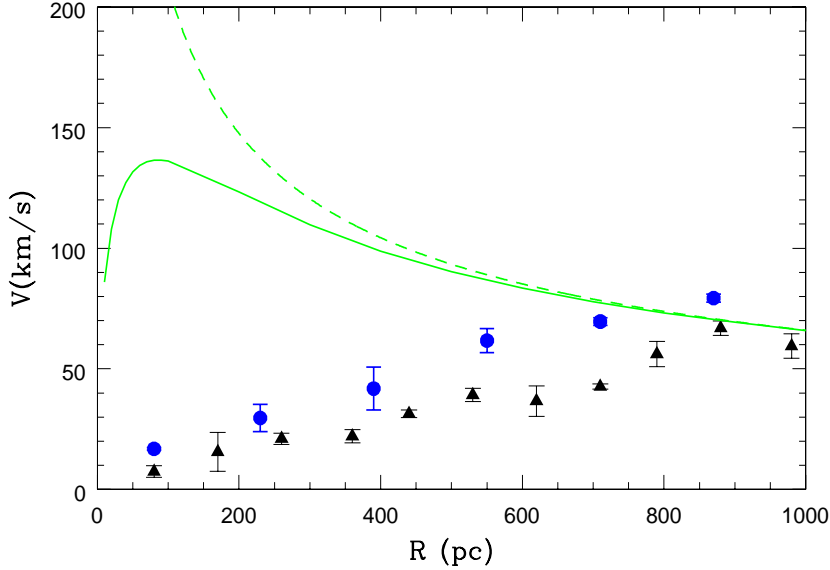}
\caption {Innermost-kpc  rotation curve of 347-g28 and 121-g6.
 We also show the circular
velocity generated by a   BH remnant of {\it average} mass  ($M_{BH}= 1  \times 10^9 \ M_\odot$)
 measured at an   infinite resolution {\it (dashed line)} and at the resolution  
 of the PS95 sample {\it (solid line).}}
\label{fig:fig1}
\end{figure}

It is worth to recall two definitions.  Given a distribution of mass
and the related gravitational potential $\phi(R,z)$, the {\it circular
velocity }$V(R)$ is defined as $V^2(R) \equiv R {d\phi(R,0)\over {dR}}$ and for 
a  point-like or a spherical distribution, the enclosed mass is  given by
$M(<R)=G^{-1}V^2 R$.  The {\it rotation curve } $V_{rot}(R)$ is
 a  set of  centered, folded and de-projected 
recessional velocities considered as a function of their galactocentric  radii. 
 These two quantities are in principle
different, in that observational errors, non-axisymmetric
disturbances, disordered motions and  a finite spatial
resolution  modify $V(R)$ into a different curve.

In this paper we assume $H_0=75 \ km/s \
Mpc^{-1}$, and we take, as the reference magnitude in the I-band,
$M_{*}=-21.9$ (which translates to $M_*=-20.5$ in the B-band
(e.g. Rhee, 1997)). No conclusion of this work depends on these  choices.

\section{QSO remnants in spirals?}

Is the available kinematical data  consistent with the centers of
spirals being the host of the silent QSO remnants, i.e.  of the
BH-powered engines that once radiated at the Eddington rate (see Paper
I)?

To introduce the data with which we will tackle this issue, we start from
the 967 rotation curve of spirals recently published in PS95, and
notice that most of them trace the kinematics down to the central kpc
(PS95, see also MFB).  These RC's have an effective  spatial resolution of
$ \simeq 1^{\prime\prime}$, i.e. $\simeq 100 {V_{sys}\over{1500 \ km/s }}\ pc
$, where $V_{sys}$ is the galaxy systemic velocity.  The range of $V_{sys}$ is
large: $600\ km/s <V_{sys}<10000 \ km/s$, so that many objects ($\sim
100$, see Appendix A) lie within a distance leading to a spatial
resolution of $\sim 100 pc$ or lesser, well sufficient to
detect a QSO remnants similar to those  commonly found in ellipticals.
 As an example, in Figure (1) we compare the innermost-kpc
region of two typical RC's belonging to our high-resolution sample (see
below) with the circular velocity  of a test particle rotating around a central body
of mass $1\times  10^9 \ M_\odot$,  corresponding  to an average 
elliptical MDO mass. In order to produce the actual rotation curve, the 
original  circular velocity  has been    line-of-sight projected, 
seeing-convolved   and  pixel averaged  (here and throughout the paper)
following  well-known  prescriptions (e.g. Appendix D in    Qian et al. 
1995).  In detail,   we
convolve $V(R,BH)$  with a $1.5''$ FWHM Gaussian seeing  and we process it by
assuming  an inclination of $60^0$, a  spatial resolution of $1.5''$
perpendicular to the slit, and of  $0.7''$ along the slit, i.e. assuming  
values   similar to those  of  the RC's of our samples (see MFB). Furthermore 
(conservatively)   $1^{\prime \prime} =100 pc$. (No conclusion  of this
paper depends on which   (reasonable)   values   are assumed   for  the above
quantities.)

 In Figure 1 we show the   
resulting   $V_{rot}(R, BH)$:   we realize that, an
average-mass, QSO powering BH, situated in the center of these
spirals, would be the major contributor to the rotation curve out to 
$\sim 1 \  kpc$,  i.e.  5-6  times their effective spatial
resolution scale.  Therefore, as we look for silent QSO in  spirals,  ground-based kinematical observations  
are  able  to detect their massive remnants, if
present.  On the other hand, the  large number of RC's in PS95/MFB sample provide 
the needed database to put this  issue on a
statistical basis.  The available data amounts to about 800 ($\times 2$, one for each arm)
measurements inside $400 \ pc$ which refer to $\sim 300$ galaxies and
four times more measurements inside  $\simeq 1 \ kpc$.  These data provide 
individual and synthetic RC's and, in turn, the circular
velocities.   
  
\begin{figure}
\vspace{8.8cm}
\includegraphics{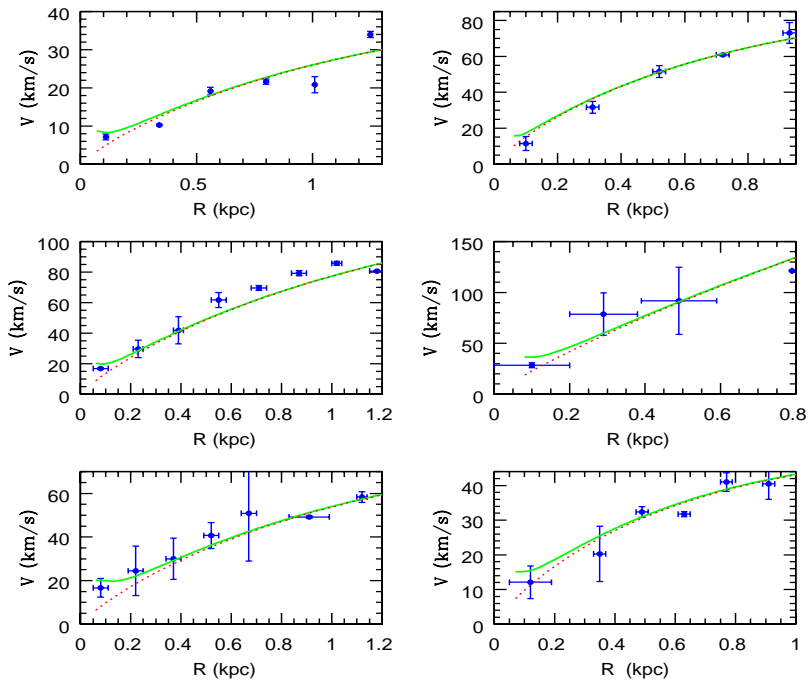}
\caption{Mass models of  rotation curves of  the IND 
sample.  The dashed line
indicates the
only-disk model while the
solid line indicates the  disk + BH model.}
\label{fig:fig1}
\end{figure}

  The data-base for our investigation includes two sub-samples
  selected from the PS95 sample.  More specifically, we select a sample,  (herafter the IND
  sample)  by including all the rotation curves that {\it individually}
   probe the mass distribution of the innermost $\sim 500 \
  pc$ and a sample (hereafter the SYN sample) by including all the
  rotation curves that {\it collectively} form  the {\it
  synthetic} rotation curve of the innermost kpc (see Rubin 1982,
  PSS).
 
The IND sample is built by selecting all PS95 RC's that
fulfill the following criteria: ({\it i)} the RC has at least 4
measurements inside a radius of $340 \ (1-{(M_I+15)\over {9}}) pc
\sim (350-500)\ pc $, {\it ii) } the innermost velocity data is
situated inside a radius of $100(1 -{( M_I+15)\over 18})\sim 
(100-150)pc$, where $M_I$ is the I-band absolute magnitude.  These
criteria are a good compromise made in order to select a large sample
of rotation curves equally distributed per magnitude interval, whilst
at the same time being able to set very stringent upper limits to the
MDO/BH mass. We find 83 RC's that result smooth, symmetric and with
negligible non-circular disturbances (see also Ratnam and Salucci,
1999) as it is also
 indicated by recent HST measurements (Erwin and  Sparke, 1999).

The SYN sample is built by selecting all the RC's in PS95 with at
least one velocity data inside a radius of $ R_{1f} =175 pc$, except
for the highest luminosity class, where we set $ R_{1f}=235 pc$.  We
found 158 RC's with (2-10) data velocities each, notice that we
neglect all the (few) velocities at distances $ \leq 25 pc$ on both
arms, because the kinematical symmetry center cannot usually be
determined with a precision better that $2 \times 25 pc =50 pc$ (see
PS95).

\section{Massive Dark Objects in Late Types Spirals}
 
To determine the upper limits for the spiral MDO masses we devise a
dual (individual/coadded) multi-step strategy: {\it 1)} we prove that
the mass distribution of a stellar disk (plus occasionally that of a
bulge) generates a circular velocity $V(R, stars)$ that reproduces very
well the observed innermost-kpc RC's and moreover, $V_{rot}(R_{res})\simeq
V(R_{res}, stars) $.  {\it 2)} we hide the biggest possible central BH
in these RC's by exploiting their r.m.s and the (small) discrepancies
between the disk mass models and the observations. {\it 3)} we correct
the derived mass upper limits for the effects of the finite spatial
resolution.

First, we derive this correction by following Qian et al, (1995). For
a point-like object observed with setup similar to that of our RC's,
to assume $V(R_{res})\equiv V_{rot}(R_{res})$ results incorrect by a
non-negligible amount $f_{res}^{1/2}$: $ V(R_{res})\simeq 1.6 V_{rot}(R_{res},BH)$
(see Fig. (1)) and therefore it leads to underestimate the MDO mass 
 by up to a factor $f_{res}\simeq 2.5$. However, if the finite resolution 
effect would dominate, we should also find: $ dV_{rot}(R, BH)/dR |_{R_{res}}\simeq
0 $.  We observe, instead: $dlog V_{rot}/dlog R |_{R_{res}} > 0.5 $
that implies that the actual values of  $f_{res}$ are  smaller than
the maximum one.  This is because one fraction of $V(R_{res})$
 originates from the diffuse stellar disk,  which is free from 
above effect, $ V(R,stars)\simeq V_{rot}(R,stars)|_{R_{res}}$
and is has a positive  log derivative 
$dlogV(R,stars)/dlogR|_{R_{res}}> 0.8$,  as the observed one.   By applying
to a disk+BH system the eqs. (1)-(5) of PS90 we  estimate that the
fraction of $V_{rot}^2(R_{res})$ due the ordinary diffuse matter is more than
$50\%$ so that: $1<f_{res}\simeq 1.6^2\times 0.5
\simeq 1.3$.  Notice that, by assuming  $f_{res}=2.5 $, no conclusion
in this paper changes.

Then, the MDO mass upper limits are computed as
$$ 
M_{MDO}\simeq (1-f)f_{res}G^{-1}V_{rot}^2(R_{res})R_{res}
\eqno(1) 
$$ 

where $1-f $ indicates the MDO-to-total mass fraction (inside
$R_{res}$) that can be hidden in the irregularities and/or in the
r.m.s. of the rotation curves,  and $f_{res}$ corrects for the
depression of the circular velocity occurring when it is measured at a
low spatial resolution.

\subsection{MDO/BH upper-limits from the IND sample}

  We reproduce the 83 RC's of the IND sample by means of two mass
  models, the first of which features: ({\it i)} a Freeman disk of
  length-scale $R_D$ (derived from photometry, PS95) and mass $M_D$
  and ({\it ii)} a MDO of mass given by Eq(1), while the second one 
 assumes  $f=1$. In Fig. (2) we show the best-fitting solution for
  the first 6 RC's of the IND sample relative to the $f=1$ (only disk, OD)
  and $0\leq f < 1$ (disk+ black hole) cases.  More details and the
  plots relative to the entire sample are given in Ratnam $\&$ Salucci
  (1999). 
 As result,  the kinematics of the crucial innermost 
   regions (i.e. $R_{res} <R
<6  R_{res}$)   never shows a  pure  CBD rotation curve, 
  neither a  RC obviously influenced by 
a  BH+resolution effects.
  On the contrary, all RC profiles strictly  conform to those generated
by a mass distribution proportional   to the light distribution,  as it is 
best recognized by noticing that, at $R\sim R_{res}$,   
  the values of the {\it local } RC slopes,
    $ dlog\ V_{rot}/dlog R = 0.85 \pm 0.1$ are in
 good agreement with those of a Freeman disk.

 Despite that the no-MDO model fits have excellent reduced 
 chi-squares: $\chi^2_\nu < 1$ and result  systematically better than the disk+ MDO model
 fits, we take the latter solutions  to determine  the MDO  mass upper
 limits.   We
 obtain values of $1-f$ that,  galaxy by galaxy,   range  between $0.2$ and  $0.7$. These  lead to the
 MDO mass upper limits plotted in  Fig (3) as a function of galaxy
 luminosity. Notice the main uncertainty  in these estimates comes
 from errors in the assumed distances that may  be as large as a factor
 2. Observational errors and mass modeling uncertainties
 induce,  in the estimated upper limits, at most   a fractional 
uncertainty of  $50\%$, that is negligible    in that  the  findings of this
paper  involve effects of the  order of one magnitude or more.  


\begin{figure}
\vspace{8.1cm}
\includegraphics{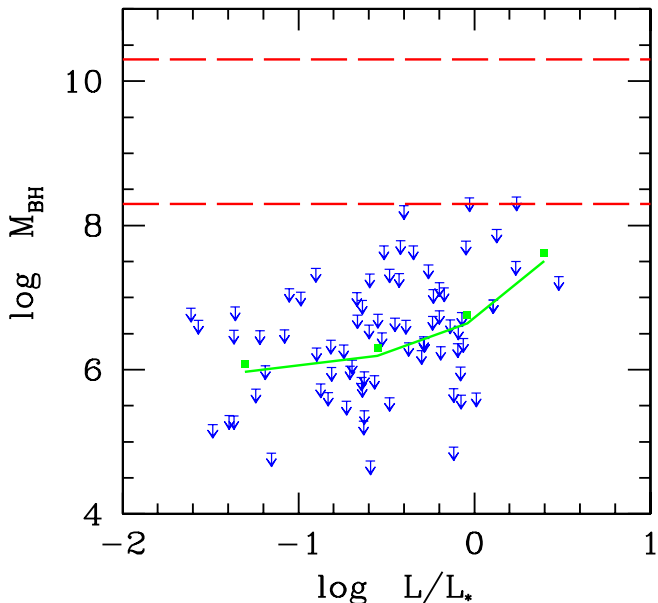}

\caption {The upper limits on the MDO mass as a function
of the luminosity for the objects of the IND sample. The solid line
represents the corresponding luminosity-averaged values. Between the dashed
lines we show  the region of   QSO  remnants.
}
\end{figure}

 
\begin{table*}
\begin{tabular}{||lcccccccc||}
\hline
\hline
$<L_I/L*>$&$<R_D>$&$\Delta$&$V_{rot }(R_{res})$&$\delta
V_{rot}(R_{res})$&$V_{rot}(R_2)$&$\delta
V_{rot}(R_2)$&$M_{disk}$&$M_b$ \\ 

(1) & (2) & (3) & (4) & (5) &(6)& (7)& (8)& (9) \\
\hline
$.05$&$1.2$&$160$&10&1.1&14.7&1.4&$1.3 \times
10^9$&0 \cr
$.19$&$2.5$&$160$&10.6&0.9&17.1&1.1& $1.0\times
10^{10}$&$0$\cr
$.48$&$3.4$&$160$&14.0&1.5&24.3&2.0&$3.0 \times
10^{10}$&$0$\cr
$1.2$&$4.3$&$160$&25.1&3.0&40.4&3.7& $1.3\times
10^{11}$&$8 \times
10^8$\cr
$3.0$&$8.7 $&$240$&73.0&10.5&78.9&14.1&$4.0
\times 10^{11}$ &$1.3
\times 10^{10}$\cr
\hline
\smallskip
\end{tabular}

\leftline{TABLE I  (1) Average luminosity for each
luminosity class 
(2) Disk length-scale  
(3) Bin size width, $pc$  ($n >1$) (4) Rotation  }

\leftline{velocity at $R_{res}$ (5) r.m.s. 
of $V_{rot}(R_{res})$ (6) Rotation velocity at $R_2$ 
(7) r.m.s. of  $V_{rot}(R_2)$ (8) Disk mass (in $M_\odot$) 
(9) Bulge}

\leftline{mass (in $M_\odot$).}  

\end{table*}

\subsection{MDO/BH upper-limits from the SYN sample}

We sort the 158 objects of the SYN sample in 5 luminosity classes of
average values $<L_I>$ (and average disk scalelenghts $<R_D>$, see
Table~1) and, for each class,  we frame the RC's with radial bins placed at the
following positions: the first bin ranges from $25 \ pc$ to $ R_{1f}$
(given in the previous section). The centers of the first bin are
placed at $R_{res}=100 \ pc$, except for the highest luminosity class where it
is placed at $R_{res}=120 \ pc$ The centers of the following bins ($n>1$)
are placed at at: $R_n= R_{1f}+(n-1) \Delta/2 $, with $\Delta(L_I/L_*)$
given in Table 1. In view of a later investigation, let us  point out that  
  the coaddition length, $>175 pc$,  for many RC's,  is larger than the 
effective spatial resolution (see Appendix A), and  
can be further  reduced for statistical tests.
\begin{figure}
\vspace{7.9cm}
\includegraphics{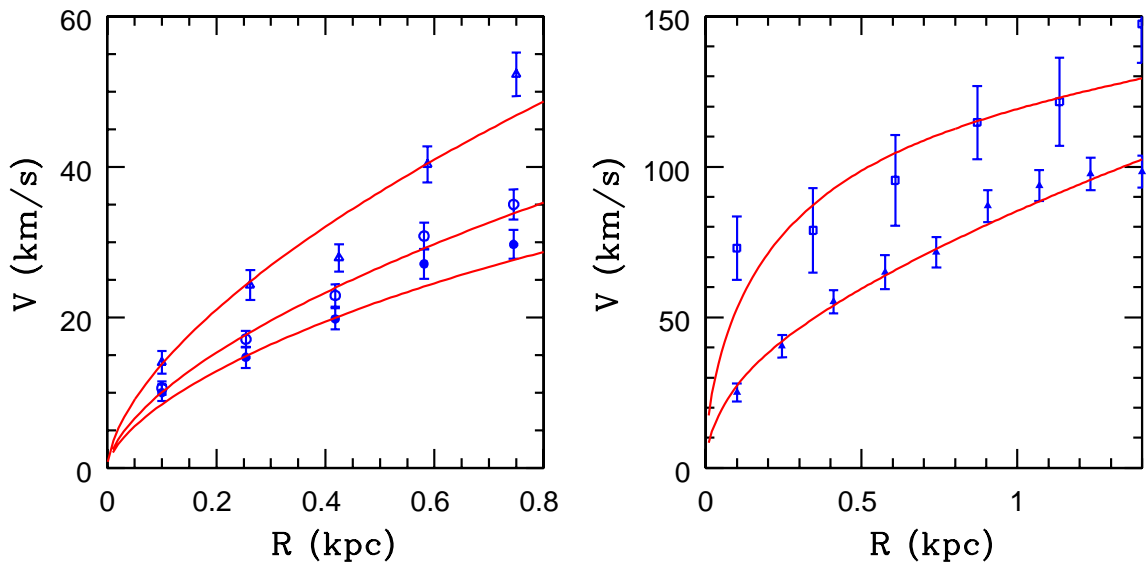}
\caption{ Synthetic innermost  rotation curves of galaxies
of different luminosities. In detail, from down to up and from left to
right, the RC correspond to objects with $<L_I/L_*>=0.05, 0.19,0.48,
1.2,3.0$, respectively. Also shown are the disk-bulge models (solid
lines). }
\end{figure}
From the rotation data relative to a radial bin (and to a luminosity
class) we derive the bin average velocity and bin  velocity r.m.s.  (i.e. 
 $V_{rot}(R_{res})$, $V_{rot}(R_2)$, $\delta V_{rot}(R_{res})$,
$\delta V_{rot}(R_2)$. The resulting synthetic RC's have 
 smooth profiles (see Fig
4) and  small  r.m.s,  $\delta V_{rot}(R_{res}) / V_{rot}(R_{res}) \sim
\delta V_{rot} (R_2)/V_{rot}(R_2 )\sim 0.1$,  indicating that  these curves  
are not significantly affected by non-circular and/or non-equilibrium
motions. Moreover, these synthetic curves 
are independent of  the coaddition scale. This is shown by means of   
 a new set of {\it zoomed } synthetic RC's that  map 
 the region out to $400 \ pc$  with  a coaddition length
half than the previous one: by comparing the two sets 
of curves we realize  that they are 
mutually consistent (see Fig 5).

  We reproduce the synthetic rotation curves by means of a mass model
  including a Freeman disk and (at high luminosities) a Hernquist
  bulge.  Disk length-scales are from PS95, bulge effective radii are
  estimated from disk-scale-lengths according to Courteau and de Jong,
  (1996). For the disk mass we adopt the values independently derived
  from the outer (1-10 kpc) kinematics (see PSS). The mass model has
  no free parameters in the three least-luminous luminosity classes,
  and one free parameter, the bulge mass, in the remaining two.  The
  mass models {\it vs} data are shown in Fig (2).  Remarkably, in the
  region presumably most affected by the presence of a central mass,
  the stellar components fully account for the observed kinematics
  (see Table 1 for the parameters of the mass models). In the
  innermost half-kpc the light traces the gravitating mass to an
  accuracy that leaves no room for a more diffuse dark component.

We   derive the MDO mass upper limits by  taking into account that 
  we are  dealing  with  coadded RC's, rather than with individual ones. Then,
      we  set up to {\it automatically}  maximise
     the  MDO 
    mass upper limits.  We assume: {\it i)} a mass model including 
    a MDO + a  stellar component with  the steepest allowed mass profile:
      $M_{stars}(R)\propto R^3 $ and,  
      {\it ii)}  the r.m.s. in  $V_{rot}(R_{res}) $ are entirely due 
      to the presence of MDO's so the upper limits derived by
      eq. (1) will be multiplied  by  the additional term
      $1+2 \delta V(R_{res}/V_{rot}(R_{res})$. 
      
      By requiring that the above mass model reproduces
      the RC's at $R_{res}$ and at  $R_2$, we get:        
 $$
f= (M(R_2)/M(R_{res})-1) ((R_2/R_{res})^3-1)^{-1}
\eqno(2)
$$ 
 where $M(R)\equiv G^{-1} V_{rot}^2R$. Specifically, $1-f \simeq 0.7$
  for  the   lowest 
 luminosity class,   continuosly rises to reach    $1-f\simeq 0.9$ at
 highest luminosity class.  
  Then,  eq. (1)-(2) and {\it ii)} determine  the (late type spirals) maximum
    central
  mass $M^u_{MDO}(L_I,Sb-Im)$ compatible with the RC's of the
   SYN sample.   Notice that the innermost-kpc 
     mass models of  objects of the IND
sample show  MDO's  masses upper limits are 
 on average   $\sim 20 \%$ smaller than those derived  
 with the above (more conservative)
assumptions.

\begin{figure}
\vspace{6.8cm}

\includegraphics{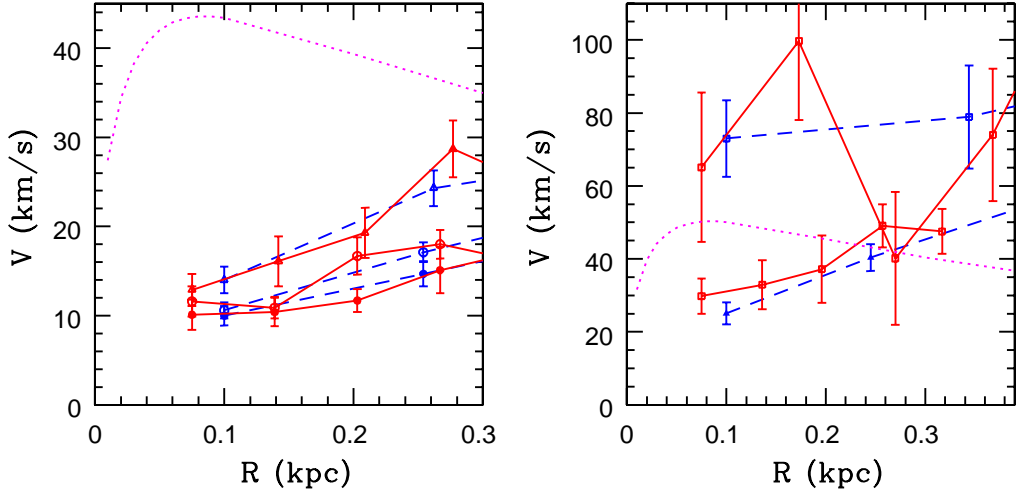}

\caption{ Coarse grained synthetic rotation curves {\it vs }
 zoomed synthetic rotation curves for galaxies of different
luminosities. Also shown the  seeing/pixel convolved RC's  of  BH's  
of $1\times 10^8 M_\odot $ (left panel) and  $3\times 10^8 M_\odot $ (right panel)
 as dotted lines. }

\end{figure}

\subsection{MDO/BH upper limits in Sb-Im}

Before we draw  our conclusion let us point to  Appendix B where we  show that  we need a very small correction
to  Eqs (1) in order to take into account the 
  part of the gravitational attraction    balanced by pressure
gradients rather than pure rotation;   more specifically. In Sb-Im's     the asymmetric drift term in the
Jeans Equation is not  a relevant fraction of the {\it circular}
velocity and therefore   the {\it rotation} velocity does not significantly  underestimate the
enclosed mass.

 The upper limits on MDO's obtained from the IND and SYN samples are
 very similar,   though the methods employed  are
 different and complementary. In detail, the "detection"  method used for the IND sample 
 has a lower  mass threshold  and it probes each
galaxy separately, while  the one employied for  the SYN sample
   is negligibly 
affected by  observational errors or galaxy peculiarities and it
involves a larger  statistics. As matter of fact, the luminosity-averaged values upper limits obtained  for the IND 
 sample (see Fig  (2)) turn out to be just $15 \%$ smaller than those   obtained 
 for  the SYN sample (shown in Fig 6).

The global  results are shown in Fig.(6): MDO's in spirals, if existing, are
remarkably less massive than those detected in ellipticals: strict
limits on their mass range between $10^6 M_\odot$ at $L_I \simeq 1/20
L_* $ to $\simeq 10^7 M_\odot$ at $\sim \ L_*$. Only in the very  most
luminous spirals (i.e. $L_I >L_*$), that amount to only few per
thousand  of the whole population,  the upper limits reach $10^8
\ M_\odot$   (barely) allowing a QSO powering MDO.
 

\begin{figure}
\vspace{8cm}
\includegraphics{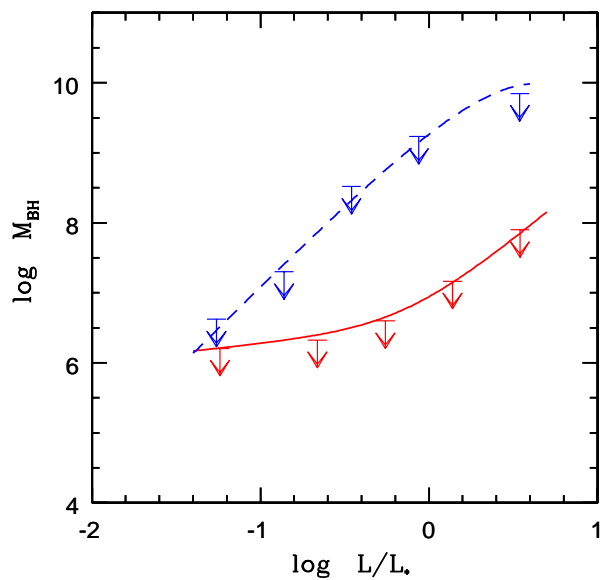}
\caption{ MDO upper limits  (arrows) and their fits as a
function of I-band luminosity for late type spirals (solid line, SYN sample) and early
type spirals (dashed-line, B-band luminosities are translated to  I-band
 luminosities following Rhee, 1997.)  }
\end{figure}

\section{MDO in Early Type Spirals}

Sa galaxies are much less numerous than late type spirals, however, in
view of their massive bulge, $M_b > 10^{10} M_\odot $, they must be
considered as a possible location for QSO remnants.  From an
observational point of view, we note that there are few available Sa
rotation curves also  because these objects have little star-formation and
weak $H_\alpha$ lines. However, the largest available sample (Rubin et
al., 1982) allows us to perform a useful investigation on the values
of MDO masses in early type spirals.  With their sample Rubin et
al. (1982) derived the synthetic (coadded) rotation curves sorting the
21 objects in 5 luminosity classes and in 8 radial bins of size $0.3
R_D$, with $R_D=2 (L_B/L_*)^{0.5} kpc$. The center of the first radial
bin is displaced from the galaxy center by $0.15 R_D$ to avoid the
complex and uncertain nuclear kinematics.  In Fig.(7) we show these
coadded curves out to $2 R_D$, and their interpolation out to $0.6
R_D$. We have checked that this interpolation out to $0.6\ R_{D}$ is
consistent with the kinematics of each individual galaxy, apart from
certain regions of complex motions, not related to the central mass.
Notice that, differently from Sb-Im galaxies, in Sa's  we are forced to
set the coaddition length much larger than the seeing or the
instrumental resolution (and  no correction for these latter is then needed).

\begin{figure}
 
\vspace{8.85cm}
\includegraphics{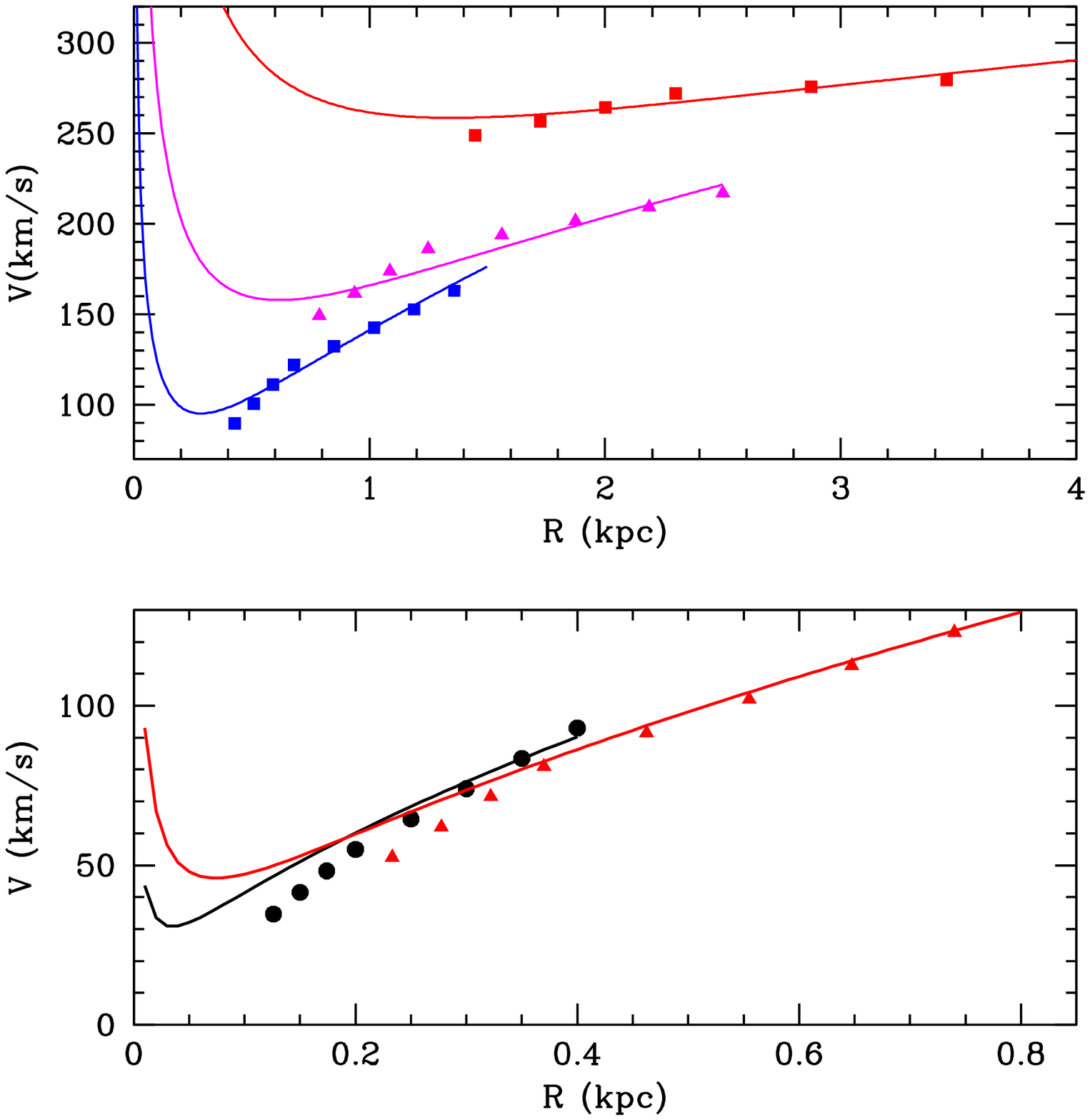}
\includegraphics{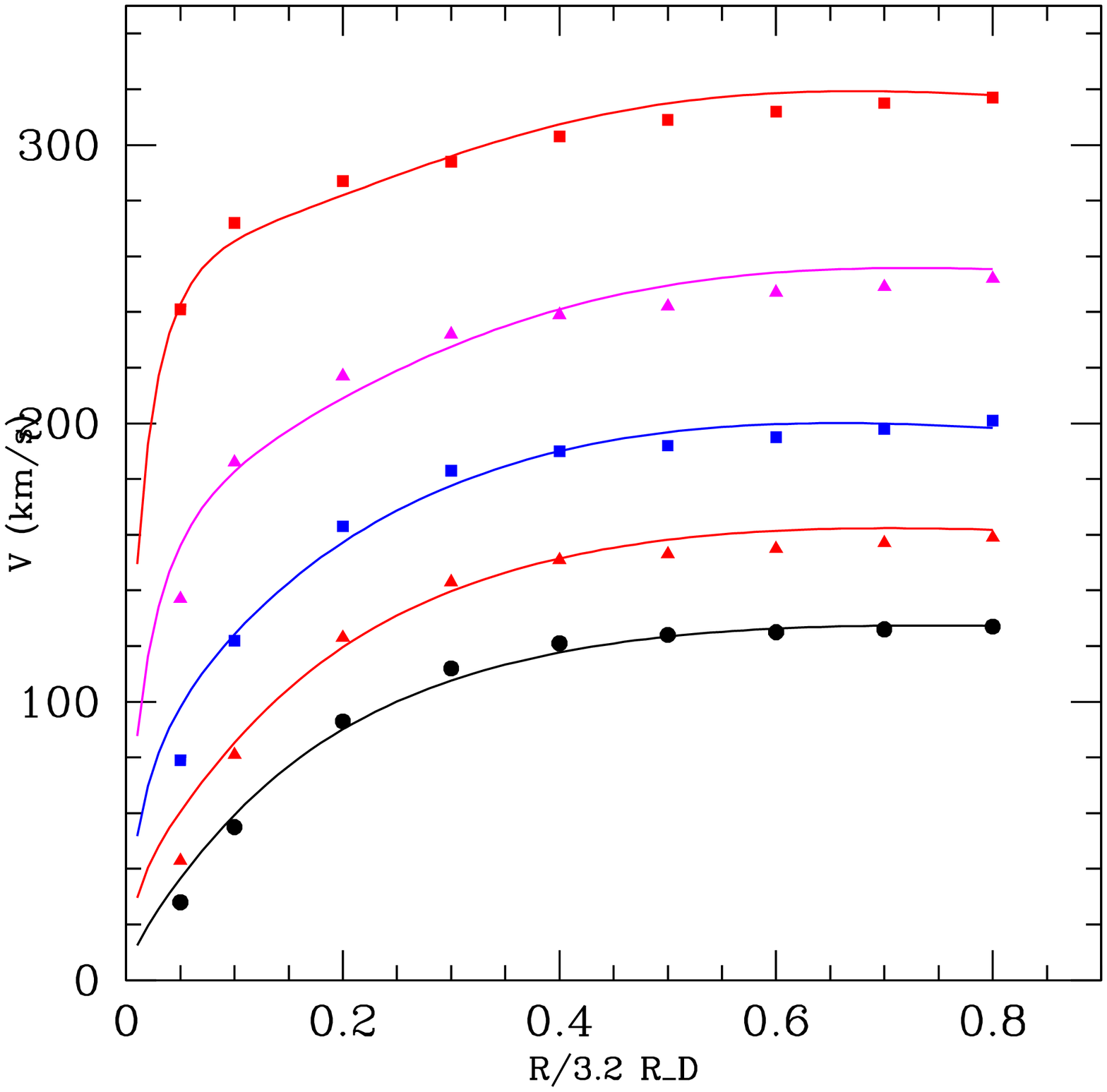}         

 \caption{ {\it Left} panel: synthetic  rotation curves of Sa
galaxies of blue magnitudes $-18, -19, -21, -22.5,-23$ and the
disk/bulge/halo best mass model.  Radial coordinate is in units of $
3.2 R_D$. {\it Right} panel: zoom of the above RC's reproduced by mass
models that include maximal MDOs (solid lines)}
\label{fig:fig4 }
\end{figure}
Sa galaxies have higher circular velocities than Sc galaxies (Rubin et
al. 1982): the total mass inside the innermost kpc is 4-10 times
larger than that in late type galaxies.  The methods devised to 
MDO upper limits   late type spirals rest upon the accurate knowledge of
the profile of the rotation curve inside the central half-$kpc$ and it
cannot be adopted in Sa's also  because the bulge-disk photometric
decomposition is uncertain. We then  devise a new method,  proper  also
to tackle a limited statistic. 

 \begin{table*}
\begin{tabular}{||lcccc||}
\hline
\hline
        $<L_B/L_*>$ &   $M_{disk}$& $M_b$& $\beta$
&$a$\\
     (1)& (2)&(3)&(4)&(5) \\
     \hline
     
        $0.06$ & $4.1 \times 10^9$ & $3.4 \times 10^7$ &
$0.49$ &
$0.2$\\           
         $0.15$ & $1.3\times 10^{10}$ & $2.2\times 10^8$
&$0.54$&$0.2$ \\
     $0.37$ & $4 \times 10^{10}$ & $3.3 \times 10^{9}$
& $0.65$ &
$0.2$\\
         $2.3$ & $1.3 \times 10^{11}$ & $2.2 \times
10^{10}$ & $0.76$
& $1.3$\\
         $3.7$ & $4 \times 10^{11} $& $7.2 \times 
10^{10}$ & $0.76$
& $1.3$\\

\hline
\smallskip
\end{tabular}
\leftline{TABLE 2. (1) $<L_B/L_*>$. (2) \& (3) Disk and
Bulge Mass
(in 
$M_\odot$)
(4) \& (5) DM parameters $\beta$ and $a$ see PSS for
details.}
\end{table*}  

  We proceed by first reproducing the coadded Sa rotation curves by
  means of no-MDO three-components mass models that include a Freeman
  disk, a Hernquist bulge and a dark halo with density profile
  according to PSS. This last component is introduced for
  completeness, but it has no role in the present analysis. We find
  that, the ordinary stellar components can very well account for the
  rotation curves out to $R<R_D$, and, by including the DM component,
  out to $2 R_D$. The parameters of the mass models, given in Table 2,
  are similar to those of Sc galaxies, with the exceptions of an
  (important) higher central density of both luminous and dark
  components and an (irrelevant) larger uncertainty on the
  distribution of DM.  We then include in the mass models a central
  mass component large enough to affect the innermost kinematics.  By
  increasing the value of the MDO mass, the models are progressively
  unable to reproduce the data until for a central mass value of
  $M_{MDO}^u(L_B,Sa)$, the model rotation curves become
  inconsistent with the coadded ones at a level $>3\delta V_{rot}$
  (with $\delta V_{rot}/V_{rot}\ \simeq 0.1$) (see Fig.(7)).

The derived MDO upper limits (see Fig.(6)) range from $5 \times 10^6
M_\odot < M_{MDO}^u< 10^{10} \ M_\odot$, and, though obtained from
measurements with high-threshold mass detection, they,  nevertheless,
indicate: 1) at a given luminosity, MDO upper limits for Sa's are up
to one order of magnitude larger than those in later Hubble types.
Apart from factors of 2-4 that could be due to  the lower spatial resolution in Sa's, this is
not a selection bias. In fact the   innermost half-kpc of early type spirals  
 contains  a larger mass than the corresponding region of  late type spirals,  and therefore 
can conceal a more massive MDO  2) 
the top $20 \%$ most luminous Sa's, $L_B>L_*$,   are serious
candidates for QSO remnants, in variance with  objects of lower luminosity,
in which the  nuclear masses do not exceed  $10^8\ M_\odot$.


\section {The MDO vs Bulge mass ratio}

Observational evidence and theoretical arguments relate the MDO mass
with that of the hot component of the host galaxy (in spirals: the
bulge).  In samples dominated by ellipticals, it has been found $log\
M_{MDO}= log M_b -2.6 \pm 0.3$ (Paper I) that is  a
proportionality between the two masses with a quite large
scatter. Here, we investigate the ratio between these two masses
  in disk systems, where
the mass of host systems is a factor 10-100 times smaller.

For Sa's the bulge masses are directly obtained by the present mass
modeling (see Table 2).  Sb-Im galaxies possess small bulges, whose
luminosity is on an average $\sim 10\%$ the stellar disk luminosity
(e.g. Simien and de Vaucouleurs, 1986).  In detail, the bulge-to-disk
luminosity ratio $k$ strongly depends on galaxy Hubble Type decreasing
from $0.25 \pm 0.05$ in Sb, to $0.03 \pm 0.01 $ in Sd's, while it is
(almost) independent of luminosity.  The bulge-to-disk mass ratio is
given by $ M_b/M_d=k (M_b/L_b)_*/(M_d/L_d)_* $ where the multiplying
term $k$ takes into account that bulges and disks have different
stellar populations and therefore different stellar mass-to-light
ratios In fact, the bulge mass-to-light ratios are higher than the
disk by a factor $\simeq 1.5$ (Van der Marel 1996; PSS). This,
inserted in the above equation, yields to two limiting relationships
between disk and bulge masses: $M_b=0.3 M_d$ and $M_b=0.05 M_d$.  For
a number of objects with known bulge mass, the MDO mass has been
directly determined (see Ho, 1998). This sample of six objects
provides the crucial calibration of the MDO-bulge systematic in
spirals.

In Fig.(8) we show the bulge-MDO mass relationship, namely with MDO
upper limits for early and late type spirals and MDO determinations
for spirals and ellipticals. Both upper limits and actual detections
show that MDOs in early type spirals may follow the elliptical
$M_{MDO}$ vs $M_b$ relationship (Paper I).  On the other hand, late
type spirals, even conceding a direct proportionality between these
masses, clearly show a lower zero-point of the MDO mass.  Notice, that
the Magorrian et al. (1998) zero-point of the MDO mass vs bulge mass
relationship implies a larger discrepancy between the values of MDO
masses in ellipticals and spirals.

 As a result, at a given bulge mass, the mass of the central dark body
 depends on the Hubble type of the host galaxy. Thus, we are lead to
 conclude that the bulge mass does not play an {\it exclusive} role in
 ``determining'' the BH mass.

\begin{figure} 
\vspace{7.5cm}
\includegraphics{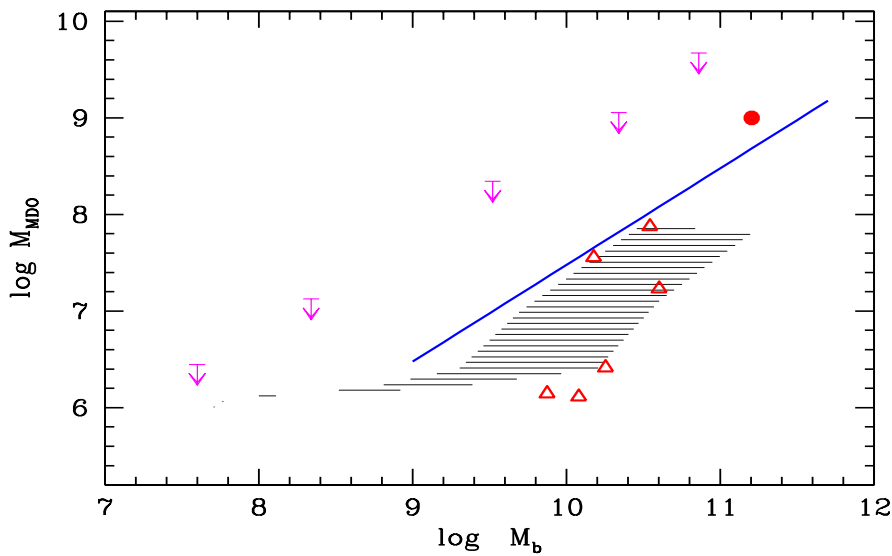}
\caption{MDO mass upper-limits  vs bulge mass for Sa
(arrows) and late type spirals (filled region) and mass detections of
 6 Sb-Im's (triangles) and one Sa (circle). The solid line represents the average
  relationship in ellipticals.}
\label{fig:fig5 }
\end{figure}

\section {Discussion and Conclusions}


     
Where are the engines of the QSO phase, i.e.  where are the masses, as
large as $\sim 10^{10} M_\odot $, of the BH remnants presently
located?  Recent observations suggest that every hot system hosts a
central BH (Magorrian et al.  1998) of mass of a few $10^{-3}$ times
the mass in stars.  Integrating the MDO upper limits over the spiral
luminosity functions we obtain an upper limit for the cosmological
density of MDOs residing in spirals. Assuming that galaxies distribute
according to Schecther luminosity functions $\phi(L_B,T)$, with the
parameters given in Paper I, we get
: $$
\rho_{BH} (Sb-Im)< 6  \times 10^4 M_\odot \ Mpc^{-3} 
$$  
$$ 
 \rho_{BH}(Sa)< 1.6 \times 10^6 M_\odot \ Mpc^{-3} 
$$
These limits must be compared with the cosmological BH mass density,
as implied by the QSO/AGN accreted material, estimated to be as large
as $ \rho_{BH} \simeq 7 \times 10^5 M_\odot \ Mpc^{-3}$ (Paper
I). Thus, the contribution to the shining phase of quasar due to BHs
hosted in late type spirals is totally negligible.
   
The role of spirals as BH hosts can be further elucidated by
determining the MDO upper limits mass function (ULMF), which can be
obtained by convolving the spiral luminosity functions with the MDO
mass limits as a function of the luminosity.  Inspection of Figure
(9) indicates that, while the QSO/AGN phenomenon implies the
existence of a large number of BH's with a wide range of masses, $\sim
10^8\ M_\odot - 2\times 10^{10} M_\odot $, the inner kinematics of
late type spirals, in agreement with the few detections available so
far, specifies that these systems can be major hosts of BHs only
for  $log M_{BH}/M_\odot < 6$. Notice that recent 
estimates  seem indeed  to  indicate
very low values for the BH masses in spirals (Colbert and
 Mushotzky, 1999)

 Thus, we can rule out
the possibility that in the distant past cold stellar systems such as
late type spirals hosted highly luminous QSOs.  The nuclei of these
objects possibly host only minor activity, of which low luminosity
Seyferts and liners represent perhaps the ``last fires in the last
galaxies''.
\begin{figure}
 
\vspace{8cm}
\includegraphics{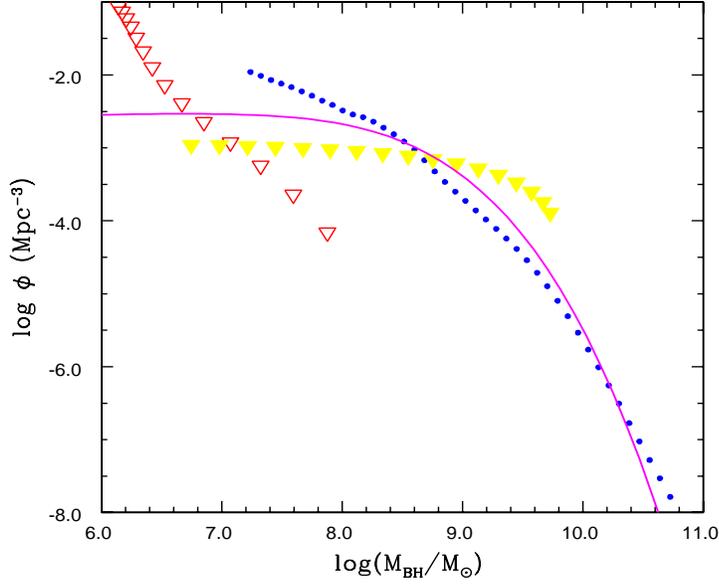}
\caption{The mass function of BH remnants (from Salucci
et al.,
1998) 
compared with the Sa and Sb-Im ULMF's. The dots
represent the
BH
remnants in QSOs, solid line the MDOs in ellipticals. 
Solid and hollow  triangles indicate ULMF's for Sa and  
Sb-Im,
respectively.}
\label{fig:fig6 }
\end{figure}
Outside these mass scales (see Fig. (9)) different Hubble types must
provide the location for the great majority of BH remnants.  In
particular, nuclei of the most luminous Sa may contribute, with
$10^{-4}\ Mpc^{-3}$ at the mass scale of $log M_{BH}/M_\odot= 8-9$, to
the locations of the present population of dormant BHs once at the
heart of the QSO shining phase. In fact, at higher BH masses the
number of Sa hosts declines exponentially, while lower masses are
outside of the QSO phenomenon.

What about for the Hubble types with peculiar luminosity and/or
central surface brightness?  Mass models of dwarf ellipticals and
rotation curves of dwarf spirals imply that in these objects the
gravitating mass inside the innermost $~200 \ pc$ is $<5 \times 10^6
M_\odot$ (Salucci \& Persic 1997); LSB rotation curves, on the other
hand, indicate that the gravitating mass within the innermost kpc is
$<5 \times 10^7 M_\odot$ (e.g.  Salucci and Persic, 1997).  In
addition, in these objects a diffuse dark matter component largely
contributes to the ``central'' dark mass. Thus, we can completely
neglect these Hubble types as host for dormant BH's in the QSO-engine
mass range.
    
A-fortiori, then, ellipticals and S0s remain the sole location for the
QSO remnants both for the common detection in the nuclei of these
objects of very massive MDOs and for the exclusion of all other galaxy
types. This conclusion is supported by recent results obtained with
HST, which show that all the QSOs brighter than $M_R=-24$ appear to  reside in elliptical
galaxies
(McLure et al.\ 1998).

\newpage

\section{Appendix A}

\begin{figure}
 
\vspace{7.6cm}
\includegraphics{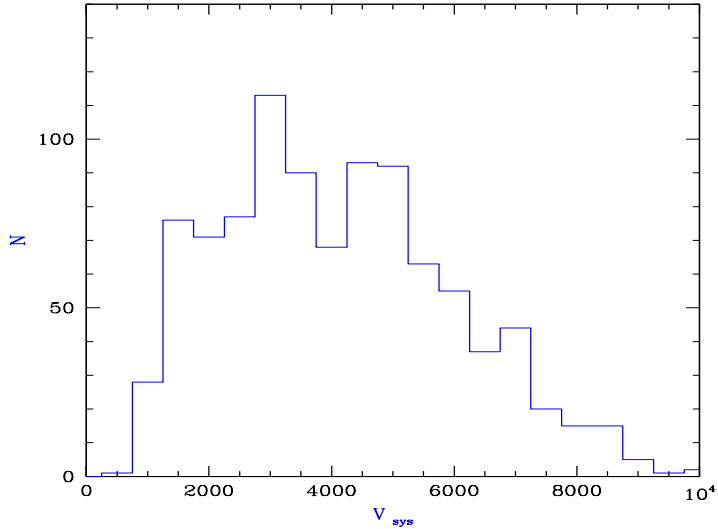}
\caption{The distribution of kinematical distances for the
PS95
sample  }
\label{fig:fig6 }
\end{figure}

In Fig (10) we show the distribution of kinematical
distances 
for the PS95 sample. The scale is:  
$1"=100 \ {V_{sys}\over{1500 \ km/s}} \ pc$. 
 
\section{Appendix B}

Altough the presence of an  axisymmetric drift,  common at very low radii in
early type galaxies (e.g. Kent 1988),   is unlikely in objects
of   later type  types  (Vega Beltran  and references therein, 1999),  it is
worth to  investigate its possible relevance on  the RC's  of  our samples.

{\it IND Sample} The OD mass model has just one free parameter,
the disk mass, determined by the outer kinematics  not
affected by non-circular motions.  Therefore, in the case of no-central MDO, 
if, at $R_{res}$,  the
axisymmetric drift  dominates,  we should detect a significant  excess of
the OD  circular velocity with respect to the observed
value $V_{rot}(R_{res})$. Since in the great majority of
the 83 RCs the OD prediction perfectly agrees with
 data,  it is unlikely that, in each galaxy, the central object plot 
 to exactly compensate  the part  of $V_{R_{res}}$ cut off by the 
 axisymmetric drift.

 {\it SYN Sample} A set of exponential thin disk RC's of different
 length-scales $R_D$ (in pc) gives rise to a set of RC's, that, when
 estimated at a {\it fixed} radius (e.g. $100 \ pc$), depend on $R_D$ as shown
 in Figure 11 (see PSS and  Salucci and Persic, 1999 for details). 
 Then, the observed  velocity-$R_D$ relationship, (see Figure 11),   
 in perfect agreement with  the above Tully-Fisher-like prediction, should be considered 
 as a further coincidence, in the case of  a systematic 
 presence of a large axisymmetric drift effect. 
   Finally, we can use the above  relationship to  
 show that, at $R_{res}$  the rotation velocity is a good measure 
 of the circular velocity:   $(V^2(R_{res})-V^2_{rot}
 (R_{res}))/V^2(R_{res})<1/3$.

\begin{figure}
\vspace{6.2 cm}
\includegraphics{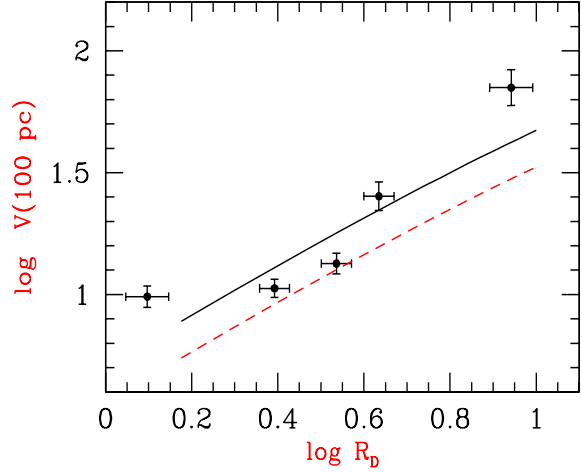}

\caption{Rotation velocity  at $R_{res}$ as a function of
disk length-scale. Data (filled circles) are compared with the
predictions of the OD  model 
 {\it (solid line)}  and those of OD +   $33\%$ non circular-motions
 model {\it (dashed line)}.  }

\label{fig:fig1}
\end{figure}

\newpage 
\medskip

\section{ACKNOWLEDGMENTS}

We thank the referee for very useful suggestions that have improved this work.
One of the authors C.R would like to thank the hospitality of
ICTP/SISSA where this work was carried out. We thank Ezio Pignatelli for 
helpful discussion. We also thank financial
support by ASI.

\bigskip
\bigskip

\centerline{REFERENCES}

\vglue 0.2truecm

{Bower G.A., etal, 1998, ApJ, 492, L111}

 Colbert, E. F.,  Mushotzky, R. F, 1999 astro-ph/9901023

 {Courteau S., Rix, H.W. 1997, AAS, 191, 7702}
 
{Courteau S., DeJong R. S., Broeils A.H., 1996, ApJ,
457, L73}

  {Cretton N., van den Bosch F.C., preprint in
astro-ph/98053324}

 Erwin, P. and Sparke, L., 1999 preprint in  astro-ph/9906262

  {Ford H. C., Tsvetanov Z.I., Ferrarese L., Jaffe W.,
1997, in Proceedings IAU Symposium 186, Kyoto, August 1997, D.B.
Sanders, J. Barnes, eds., Kluwer Academic Pub.}

   {Ho L.C., 1998, in Observational
Evidence for Black Holes in the Universe, ed. Chakrabarti, S. K.,
Kluwer Academic Pub.}
 
   {Kormendy J., Richstone D., 1995, ARAA, 33, 581}

   {Magorrian J., Tremaine S., Richstone D., Bender R.,
Bower G., Dressler A., Faber S.M., Gebhardt K., Green R.,
Grillmair C., Kormendy J., Lauer T.R., 1998,  115, 2285}

Matheweson, D.S, Ford, V.L., Buchorn, M. 1992, ApJS, 81,413,  MFB

  { McLure R. J., Dunlop J. S. , Kukula M.J, Baum S.
A., O'Dea C.  P., Hughes D. H. , 1998, preprint in astro-ph 9809030}

   {Persic M., Salucci P., 1990, MNRAS, 247, 349 PS90}

{Persic M., Salucci P., 1995, ApJS, 99, 501}

   {Persic M., Salucci P., Stel F., 1996, MNRAS, 281, 27, PSS}

Qian, E.E., de Zeew, P.T., van der Marel R.P., Hunter, C. 1995, MNRAS, 274, 602 

   {Rhee M.H. 1997, Phd. Thesis, Groningen University}

   {Ratnam C. , Salucci P. 1999,  submitted }

  {Rubin V.C., Thonnard N., Ford W.K, Burstein D,
1982, ApJ, 261,439}

   {Salucci P. , Persic M. ASP Conference Series 117, 1 ,
1997 Eds: Persic. M and Salucci P. }  

  {Salucci P., Szuskiewicz E., Monaco P., Danese G.,
1999 MNRAS, in press, Paper I }

  {Schreier E.J. etal preprint,  astro-ph/9804098}

  {Simien F., DeVaucouleurs G., 1986, ApJ, 302, 564}

   {van der Marel R.P., 1997,
in Proceedings IAU Symposium 186, Kyoto, August 1997, D.B.  Sanders,
J. Barnes, eds., Kluwer Academic Pub.}

   {van der Marel R.P., 1998,   astro-ph 9806365 }

Vega Beltran, J.C., Zeilinger, W.W., Pizzella A., Corsini, E.M., Bertola, F., Beckman, J.,  1999, astro-ph 9906349

\vfill\eject

\end{document}